\documentclass[aps,twocolumn,amssymb,superscriptaddress]{revtex4-1}
\usepackage[final]{changes}
\usepackage[colorlinks=true,citecolor=blue,linkcolor=blue,urlcolor=blue]{hyperref}
\usepackage{amsmath}
\usepackage{amssymb}
\usepackage{bbm}
\usepackage{xcolor}
\usepackage{graphicx}
\usepackage{dcolumn}
\usepackage{bm}
\usepackage{hyperref}
\usepackage{afterpage}
\usepackage{tabularx}
\graphicspath{{Latexpicture/}}

\begin{document}
\title{Deterministic two-photon C-Z gate with the two-photon quantum Rabi model}

\author{Jia-Cheng Tang}
\affiliation{Hunan Key Laboratory for Micro-Nano Energy Materials and Devices\\ and School of Physics and Optoelectronics, Xiangtan University, Hunan 411105, China}

\author{Jin Zhao}
\email{zhaojin24@mails.ucas.ac.cn}
\affiliation{Hangzhou Institute for Advanced Study, UCAS,
Hangzhou 310024, China}
\author{Haitao Yang}
\affiliation{School of physics and electronic information engineering, Zhaotong University,
 Zhaotong 657000, People's Republic of China}

\author{Junlong Tian}
\affiliation{Department of Electronic Science, College of Big Data and Information Engineering, Guizhou University, Guiyang 550025, China}
\author{Pinghua Tang}
\affiliation{Hunan Key Laboratory for Micro-Nano Energy Materials and Devices\\ and School of Physics and Optoelectronics, Xiangtan University, Hunan 411105, China}
\author{Shuai-Peng Wang}
\email{wangsp@baqis.ac.cn}
\affiliation{Beijing Academy of Quantum Information Sciences, Beijing 100193, China}

\author{Lucas Lamata}
\email{llamata@us.es}
\affiliation{Departamento de F\'isica At\'omica, Molecular y Nuclear, Universidad de Sevilla, 41080 Sevilla, Spain}

\author{Jie Peng}
\email{jpeng@xtu.edu.cn}
\affiliation{Hunan Key Laboratory for Micro-Nano Energy Materials and Devices\\ and School of Physics and Optoelectronics, Xiangtan University, Hunan 411105, China}
\begin{abstract}
We propose a scheme for realizing a deterministic two-photon C-Z gate based on variants of the two-photon quantum Rabi model (QRM), which is feasible within the framework of circuit QED. We begin by utilizing the two-photon interaction to implement the nonlinear sign (NS) gate, and subsequently, we construct the C-Z gate following the KLM scheme. We consider three different regimes: the strong coupling regime, the perturbative ultrastrong coupling regime, and the large detuning regime. Our results indicate that the C-Z gate operates fast with high fidelity, and is robust against decoherence. We also show the photonic state in the waveguide can be input into the circuit QED system through a variable coupler, and released after interaction with almost the same waveform except for a $\pi$-phase shift. Our scheme offers a suitable approach for achieving fast and deterministic two-photon quantum gates via light-matter interactions.
\end{abstract}

\maketitle
\section{Introduction}\label{I}

Quantum computing offers advantages in certain fields due to its unique computational methods \cite{qca}. The power of quantum computing grows exponentially with the number of qubits involved in the computation, a feature that stems from the quantum properties of superposition and entanglement \cite{qmh,qc}. A variety of physical systems have been explored for implementing quantum computing, including photons \cite{asf,pqt,oqc,hfp}, trapped ions \cite{afa}, nuclear magnetic resonance \cite{bsr}, quantum dots \cite{cqc}, cavity QED \cite{spq}, and circuit QED \cite{aqe,mpw}. Among these, superconducting circuit QED systems \cite{qsu,sqc} and photonic quantum computation \cite{qcau} have garnered significant attention due to substantial experimental and theoretical progress over the past two decades \cite{ciq,sqo,cqe}.

Universal quantum gates, which are essential components of a universal quantum computer, can be constructed using C-Z gates (or C-phase gates) in conjunction with a series of single-qubit rotations \cite{qca}. Compared to circuit QED systems, implementing a C-Z gate in a photonic system is more challenging due to the scarcity of nonlinearity. In 2001, Knill, Laflamme, and Milburn (KLM) proposed a complete linear optics quantum computing scheme \cite{asf}. They demonstrated that linear optics elements could be used to implement a nonlinear sign gate (NS gate), and a C-Z gate could be constructed using two NS gates. However, the NS gate and C-Z gate in their scheme are nondeterministic, with success probabilities of 1/4 and 1/16, respectively, and they require extensive optical resources for scalability \cite{oqc}. A significant challenge in constructing two-qubit logical gates in optical quantum computing is the introduction of optical nonlinearity. In the KLM scheme, the necessary nonlinearity is achieved through single-photon detection and postselection.
Alternative methods for implementing NS gates and, consequently, C-Z gates have been proposed. An early approach involved Kerr-nonlinear photonic crystals \cite{qcw}. In 2017, Costanzo et al. emulated the effect of strong Kerr nonlinearity on the quantum state of light using sequences and superpositions of single-photon additions and subtractions \cite{mis}. Others have introduced nonlinearity through photon-atom interactions \cite{ict,qno,apq}. In 2013, Adhikari et al. achieved a deterministic two-photon nonlinear phase shift in the microwave domain using a nonlinearly-coupled system between an LC resonator and a transmon qubit via an adiabatic scheme \cite{noq}. Another deterministic C-Z gate scheme, which exploited a nonlinear $\pi$-phase shift of two-photon molecule generations in chiral waveguide QED systems, was proposed in 2019 \cite{dtp}. Additionally, several photonic quantum gates have been proposed in recent years \cite{tcg,pfp,htq}.

In this article, we propose a scheme to realize a deterministic two-photon C-Z gate based on the two-photon quantum Rabi model (QRM). The QRM is extensively studied recently \cite{iot,chen1,lv1,ying,lizimin,larson,oso,dets}. The two-photon QRM can be directly implemented in a circuit QED system with a sophisticated design by Felicetti et al. in 2018 \cite{tpq,ucr}, where a flux qubit is inductively
coupled to a SQUID. An effective two-photon QRM with arbitrary coupling
 strength can also be realized in circuit QED by adding a two-tone drive \cite{ayysha}. This two-photon interaction reveals novel physical phenomena \cite{sot,eso,scv,puc} and can be directly applied to introduce a two-photon nonlinear phase shift, thereby enabling the implementation of the NS gate. We consider three scenarios: first, the system is described by the two-photon Jaynes-Cummings (JC) model in the strong coupling (SC) regime; second, a Bloch-Siegert Hamiltonian is applied in the perturbative ultrastrong coupling (p-USC) regime; and third, a dispersive Hamiltonian is used to describe the system in the large detuning regime. We calculate the dynamics and find that a $\pi$-phase can be accumulated for the two-photon state, allowing for the realization of a two-photon NS gate in 1.4, 0.8, and 15 ns with high fidelity, respectively. Considering the full experimental process, we show the photonic states in the waveguide can be input into the cavity through a variable coupler, and released after interaction with the flux qubit. The waveform is almost kept the same, but a $\pi$-phase shift is added.  We then use this to construct the two-photon C-Z gate according to the KLM scheme \cite{asf}. Furthermore, we demonstrate that our scheme is not only fast but also robust against decoherence \cite{uqg}.

\section{Two-photon C-Z gate in the strong coupling regime}\label{II}
\label{sec.II}

\begin{figure}[h]
\centering
\includegraphics[scale=0.35]{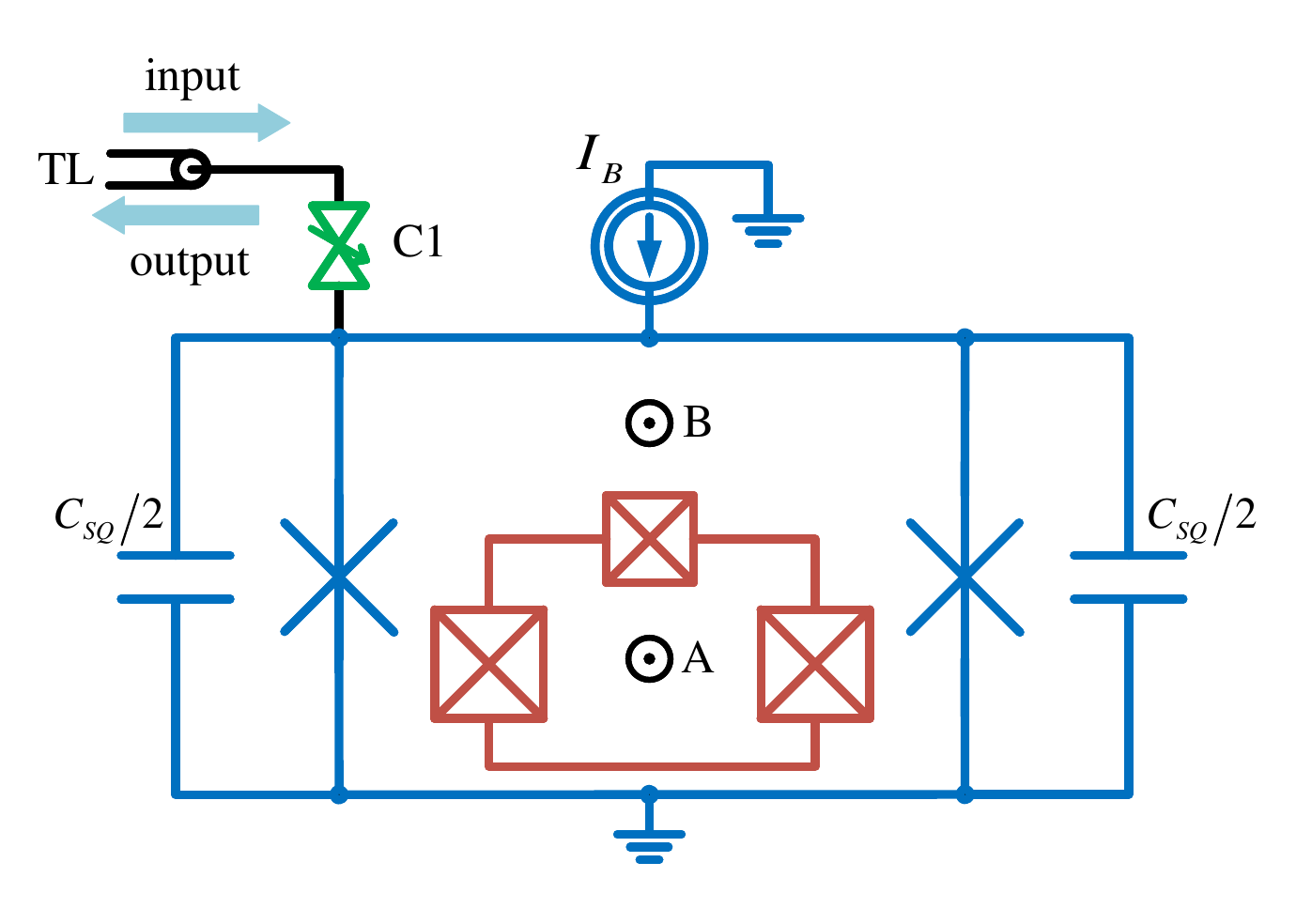}
\renewcommand\figurename{\textbf{FIG.}}
\caption[1]{Schematic diagram of the circuit QED setup to implement a NS gate. A three-junction flux qubit (red) is inductively coupled to a SQUID resonator (blue), which is biased by an external direct current $I_B$ \cite{tpq}. The SQUID is connected to an input/output transmission line (TL) via a variable coupler (green), enabling catching and releasing of photons with a controllable photon dissipation rate $\kappa_c$ \cite{car}. An external flux $\varPhi_{\text{A}}$ threads the flux qubit in region A, and the larger loop of the SQUID outside region A is threaded by a flux $\varPhi_{\text{B}}$. Thus, the net external flux through the SQUID is $\varPhi_{\text{ext}}=\varPhi_{\text{B}}+\varPhi_{\text{A}}$.}
\label{fig.1}
\end{figure}

The NS gate is a critical component of the C-Z gate within the KLM scheme \cite{asf}. It introduces a nonlinear $\pi$-phase shift specific to the two-photon component, which can be represented as the transformation $\alpha_0\vert0\rangle+\alpha_1\vert1\rangle+\alpha_2\vert2\rangle \rightarrow \alpha_0\vert0\rangle+\alpha_1\vert1\rangle-\alpha_2\vert2\rangle$. This effect is achieved through a nonlinear interaction between a qubit and a photonic mode. Typically, linear coupling prevails, but a two-photon quantum Rabi model can be implemented in circuit QED with a sophisticated design by Felicetti et al. \cite{tpq}. This design involves a superconducting quantum interference device (SQUID) inductively coupled to a three-junction flux qubit. By connecting it to an input/output transmission line (TL) via a variable coupler, the realization of a two-photon NS gate becomes feasible, as depicted in Fig. \ref{fig.1}.

The dynamics of the two-photon quantum Rabi model (QRM) encompass infinite-photon subspaces, complicating the direct realization of a two-photon gate. However, we can address this challenge within three specific regimes: the strong coupling (SC) regime, the perturbative ultrastrong coupling (p-USC) regime, and the large detuning regime. Here we focus on the first case, where the Hamiltonian can be effectively approximated by the two-photon Jaynes-Cummings (JC) Hamiltonian under the rotating-wave approximation ($\hbar=1$)
\begin{eqnarray}
H_{2JC}=\omega_r a^\dagger a+\frac{\omega_q}{2}\sigma_z+g(\sigma_+a^{2}+\sigma_-a^{\dagger2}),
\end{eqnarray}
where $\omega_r=\sqrt{(C_{\text{SQ}}\varPhi_0)/(4\pi I_c \text{cos}(\pi\varPhi_{\text{ext}}/\varPhi_0))}$. $C_{\text{SQ}}$ is the total capacitance of the SQUID and $I_{c}$ is the critical current of the single Josephson junction. $\varPhi_{0}=h/(2e)$ is magnetic flux quantum. $\varPhi_{\text{ext}}$ is the total magnetic flux threading the SQUID loop. $a^\dagger (a)$ is the photonic creation (annihilation) operator. $\omega_q$ is the transition energy of two lowest persistent-current states of the flux qubit which has been biased at the degeneracy point $\varPhi_{\text{A}}=\varPhi_0/2$ \cite{jpc}. $\sigma_z=\vert e\rangle\langle e\vert-\vert g\rangle\langle g\vert$, with $\vert e\rangle$ and $\vert g\rangle$ denoting the left-circulating and right-circulating persistent-current states, respectively. $\sigma_+=\vert e\rangle\langle g\vert$ and $\sigma_-=\vert g\rangle\langle e\vert$ are the raising and lowing operation of the flux qubit respectively. The coupling strength between the SQUID and the flux qubit $g=-(\pi/4)\text{tan}(\pi\varPhi_{\text{ext}}/\varPhi_0)(MI_p/\varPhi_0)\omega_r$, where $I_{p}$ is the persisting current of the flux qubit and $M$ is the coefficient of mutual inductance between two loops. 
\begin{figure}[b]
\centering
\includegraphics[scale=0.3]{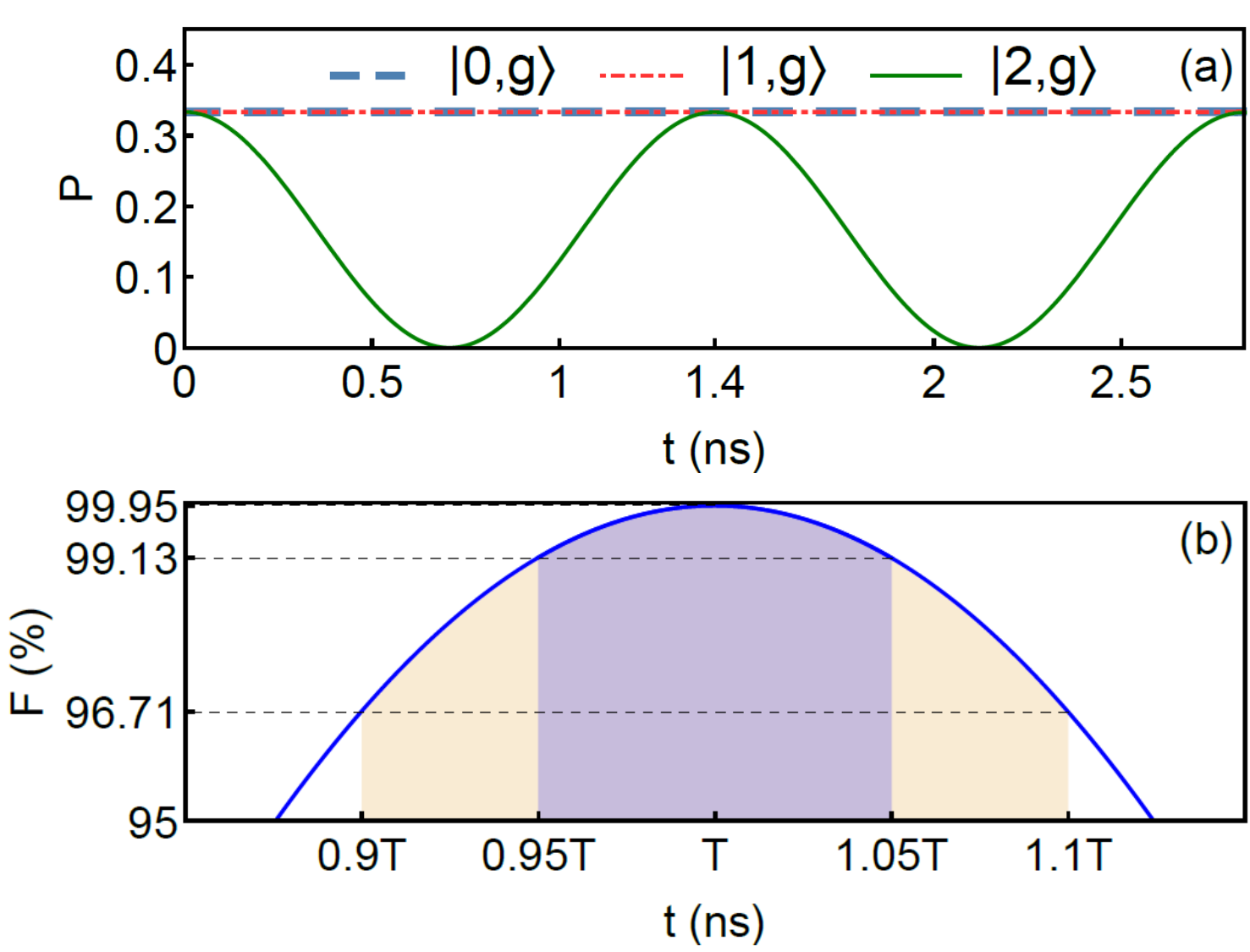}
\renewcommand\figurename{\textbf{FIG.}}
\caption[1]{(a) Time evolution governed by $H_{2JC}^{int}$ with initial state $\vert\psi_{0}\rangle$ ($\alpha_0=\alpha_1=\alpha_2$ for simplicity). (b) Fidelity of the NS gate near the optimum time $T=1.4\:\text{ns}$. $\omega_{r}/2\pi=\omega_{q}/4\pi=5\:\text{GHz}$, $g/2\pi=0.25\:\text{GHz}$, and environment dissipations are $\kappa=\gamma=\gamma_{\varphi}=0.05\:\mu\text{s}^{-1}$.}
\label{fig.2}
\end{figure}
\begin{figure}[b]
\centering
\includegraphics[scale=0.3]{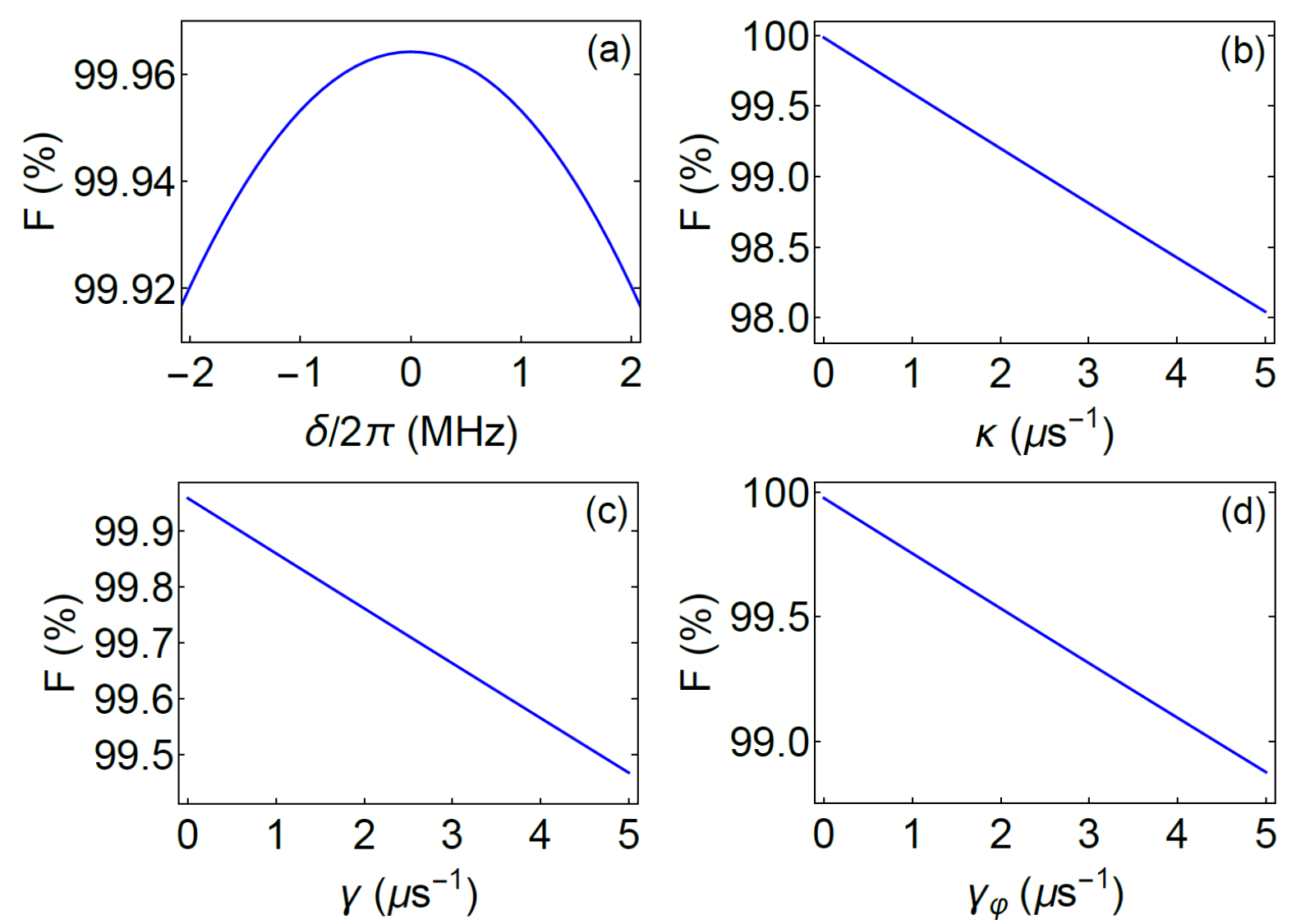}
\renewcommand\figurename{\textbf{FIG.}}
\caption[1]{The impact of detuning and environment dissipations on the fidelity $F$ of the NS gate. (a) $F$ as a function of the detuning $\delta$ between the photon and qubit frequencies, where $\kappa=\gamma=\gamma_\varphi=0.05\:\mu\text{s}^{-1}$. (b) $F$ as a function of the resonator dissipation rate, where $\delta/2\pi=1\:\text{MHz}, \gamma=\gamma_\varphi=0.05\:\mu\text{s}^{-1}$. (c) $F$ as a function of the energy relaxation rate $\gamma$ of the qubit, where $\delta/2\pi=1\:\text{MHz}, \kappa=\gamma_\varphi=0.05\:\mu\text{s}^{-1}$. (d) $F$ as a function of the dephasing rate of the qubit, where $\delta/2\pi=1\:\text{MHz}, \kappa=\gamma=0.05\:\mu\text{s}^{-1}$.}
\label{fig.3} 
\end{figure}

In the two-photon JC model, it is easy to find a symmetry operator $C_{2JC}=a^{\dagger}a+2\sigma_+\sigma_-$ commuting with $H_{2JC}$. The symmetry leads to exact solutions in subspaces spanned over the basis $\lbrace\vert0,g\rangle\rbrace,\lbrace\vert1,g\rangle\rbrace,\lbrace\vert2,g\rangle,\vert0,e\rangle\rbrace\\...\lbrace\vert n,g\rangle,\vert n-2,e\rangle\rbrace$. Transformed to the interaction picture with respect to $H_0=\omega_{r}a^{\dagger}a+\omega_{q}/2\;\sigma_{z}$, the Hamiltonian of the two-photon JC model reads
\begin{eqnarray}
H_{2JC}^{int}=g(\sigma_+a^{2}e^{i\delta t}+\sigma_-a^{\dagger2}e^{-i\delta t}).
\end{eqnarray}
Here, $\delta=\omega_q-2\omega_r$ is the detuning between the qubit and two photons. We consider the resonant case and set $\omega_q=2\omega_r$ accordingly.
Suppose the input microwave photon state from the TL is an arbitrary superposition state $\alpha_0\vert0\rangle+\alpha_1\vert1\rangle+\alpha_2\vert2\rangle$ 
and the flux qubit has been cooled down to the ground state. First, we turn on the variable coupler C to input the initial state
\begin{align}
\vert\psi_0\rangle=(\alpha_0\vert0\rangle+\alpha_1\vert1\rangle+\alpha_2\vert2\rangle)\otimes\vert g\rangle.
\label{eq.psi0}
\end{align}
The eigenstates $\vert0,g\rangle$ and $\vert1,g\rangle$, belonging to the zero-excitation and one-excitation subspaces respectively, are time-invariant. But in the two-excitation subspace $\lbrace\vert2,g\rangle,\vert0,e\rangle\rbrace$, the system will evolve in the form of Rabi oscillation, as shown in Fig.\:\:\hyperref[fig.1]{2(a)}. By solving the Schr\"odinger equation, we obtain
\begin{align}
\vert\psi(t)\rangle=&\alpha_0\vert0,g\rangle+\alpha_1\vert1,g\rangle\nonumber\\
&+\alpha_2\big(\cos(\sqrt{2}gt)\vert2,g\rangle-i\sin(\sqrt{2}gt)\vert0,e\rangle\big).
\label{psit_2JC}
\end{align}

After resonant interaction with an operation time of $t=\pi/(\sqrt{2}g)$, the system will evolve into 
\begin{eqnarray}
\psi_{\text{ideal}}=\vert\psi(\frac{\pi}{\sqrt{2}g})\rangle=\alpha_0\vert0,g\rangle+\alpha_1\vert1,g\rangle-\alpha_2\vert2,g\rangle
\end{eqnarray}
Then we turn on  C to release photons from the resonator into the TL as an output.

\begin{figure}[t]
\centering
\includegraphics[scale=0.3]{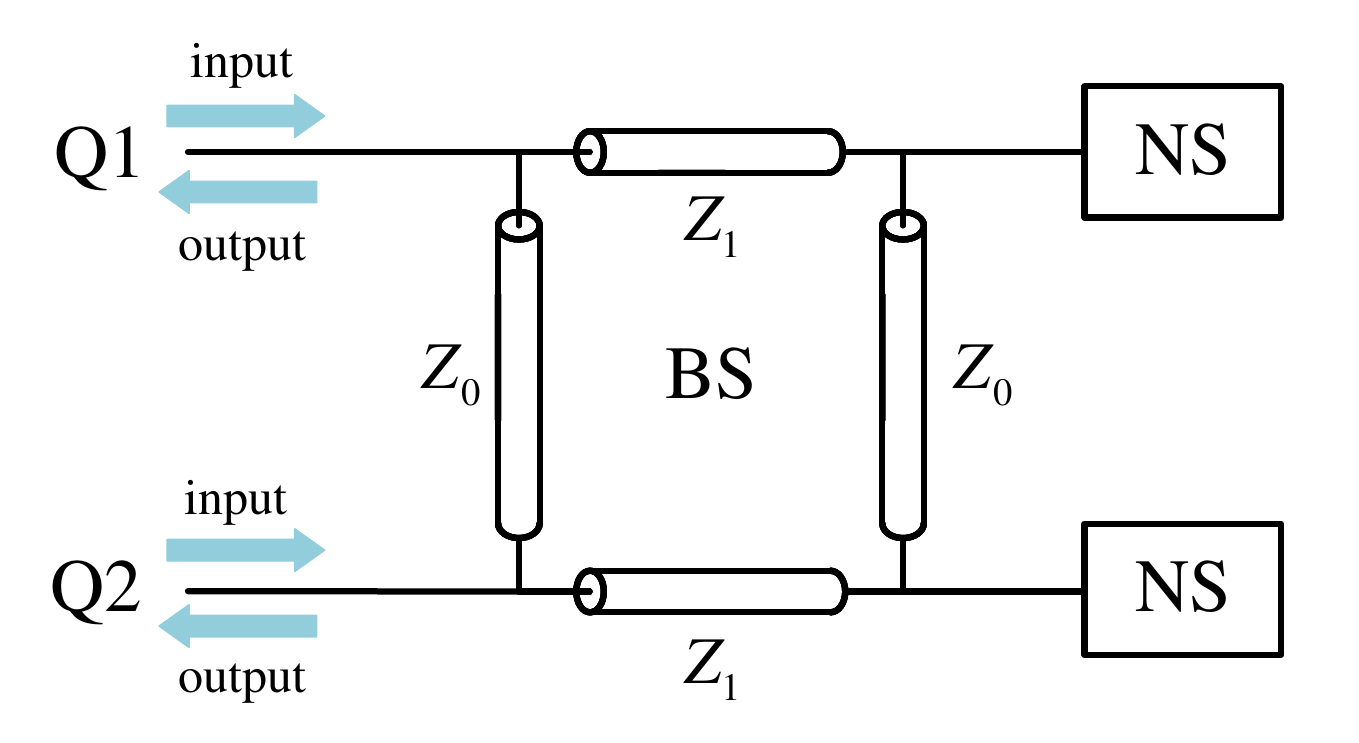}
\renewcommand\figurename{\textbf{FIG.}}
\caption[1]{Schematic diagram of two-photon C-Z gate. Four $\lambda/4$ sections of TL with impedance $Z_0$ and $Z_1=Z_0/\sqrt{2}$ realize a 50:50 beam-splitter \cite{aom,me}. Q1, Q2 act as both input and output ports of the C-Z gate. The NS gate is realized according to the design in Fig. \ref{fig.1}}
\label{fig.4}
\end{figure}

To evaluate the performance of the NS gate, we simulate the evolution of the system under environment dissipations with the Lindblad master equation \cite{cqe}
\begin{small}
\begin{eqnarray}
\dot\rho(t)=-i[H_{2JC}^{int},\rho]+\kappa D[a]\rho+\gamma D[\sigma_-]\rho+\frac{\gamma_\varphi}{2}D[\sigma_z]\rho,
\label{eq.drhodt}
\end{eqnarray}
\end{small}
where $D[L]\rho=(2L\rho L^{\dagger}-L^{\dagger}L\rho-\rho L^{\dagger}L)/2$ with $L=a,\sigma_-,\sigma_z$. $\kappa$ is the dissipation rate of the SQUID resonator. $\gamma$ is the energy relaxation rate of the qubit, and $\gamma_\varphi$ is the dephasing rate of the qubit. The parameters are chosen as: $\omega_r/2\pi=\omega_q/4\pi=5\:\text{GHz}$, $g/2\pi=0.25\:\text{GHz}$, and $\kappa=\gamma=\gamma_\varphi=0.05\:\mu\text{s}^{-1}$ \cite{cqe}. The simulation results in Fig. \hyperref[fig.2]{2(b)} shows that the fidelity of our NS gate $F=\vert\langle\psi_{\text{ideal}}\vert\rho(T)\vert\psi_{\text{ideal}}\rangle\vert$ is 99.95\% within the operation time of $T=2\pi/g\approx 1.4$\:\text{ns}, which can be further reduced to 0.7\:\text{ns} at the SC limit ($g=0.1\omega_{r}$). It is worth noting that the coherence time of flux qubit has reached 100 $\mu\text{s}$\cite{aqe,uff}, and the resonator lifetime has been extended to milliseconds \cite{neo}. We have also investigated the effect of deviations in the evolution time. Even with a $\pm5\%$ deviation from the optimal point $T$, the fidelity remains above 99.13\%, as depicted in Fig. \hyperref[fig.2]{2(b)}. The influences of detuning and environment dissipations on fidelity are shown in Fig. \ref{fig.3}. It can be seen that the fidelity is robust against these factors, although it is more sensitive to $\kappa$ and $\gamma_{\varphi}$ than to $\gamma$.

\begin{figure*}[t]
\centering
\begin{minipage}{0.43\textwidth}
\includegraphics[width=\linewidth]{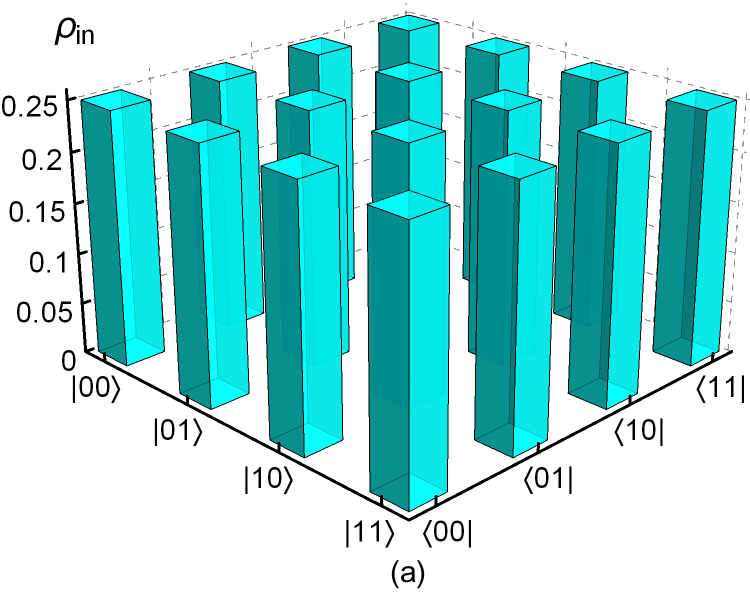}
\end{minipage}
\hfill
\begin{minipage}{0.06\textwidth}
\includegraphics[width=\linewidth]{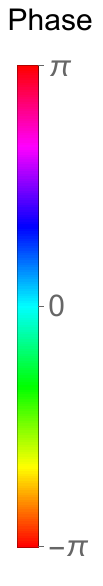}
\end{minipage}
\hfill
\begin{minipage}{0.43\textwidth}
\includegraphics[width=\linewidth]{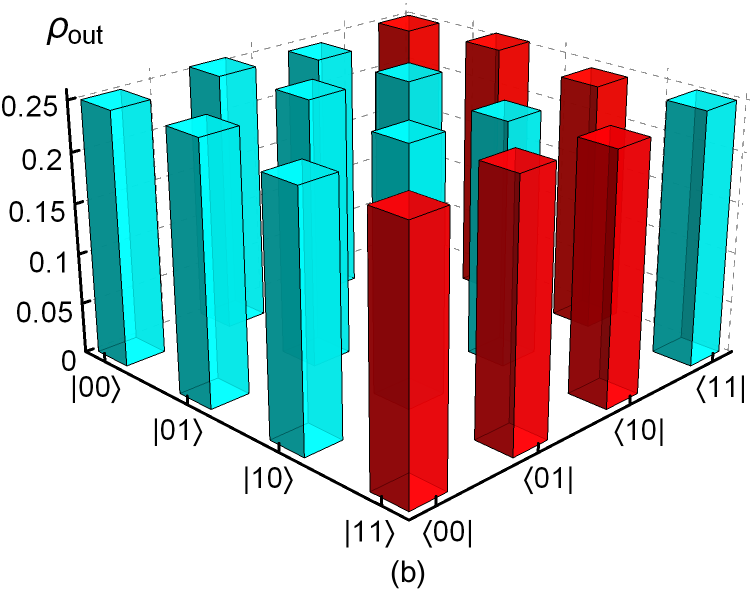}
\end{minipage}
\renewcommand\figurename{\textbf{FIG.}}
\caption[1]{(a) Density matrix $\rho_{\text{in}}$ of input state $\vert\psi_{\text{in}}\rangle=\vert00\rangle+\vert01\rangle+\vert10\rangle+\vert11\rangle$. (b) Density matrix $\rho_{\text{out}}$. Here, we choose the environment parameters $\gamma=\kappa=\gamma_\varphi=0.05\:\mu\text{s}^{-1}$, and the beam-splitter parameter $\theta=\pi/4+0.01$.}
\label{fig.5}
\end{figure*}
Next we construct the two-photon C-Z gate with the NS gate designed above, according to the KLM scheme \cite{asf} (Fig. \ref{fig.4}). Quantum information is encoded on a two-rail photon qubit basis $\lbrace\vert0\rangle_1\vert0\rangle_2$, $\vert0\rangle_1\vert1\rangle_2, \vert1\rangle_1\vert0\rangle_2, \vert1\rangle_1\vert1\rangle_2\rbrace$. Since the total photon number is conserved, the involved photonic states include $\lbrace\vert00\rangle, \vert01\rangle, \vert10\rangle, \vert02\rangle, \vert11\rangle, \vert20\rangle\rbrace$. The beam-splitters (BS) used to separate and mix microwave light \cite{rbi} can be described by an unitary matrix
\begin{align}
&U(\theta)=\nonumber\\
&\resizebox{0.48\textwidth}{!}{
$\begin{pmatrix}
1 & 0 & 0 & 0 & 0 & 0\\
0 & -\sin\theta & \cos\theta & 0 & 0 & 0\\
0 & \cos\theta & \sin\theta & 0 & 0 & 0\\
0 & 0 & 0 & \sin^2\theta & -\sqrt{2}\sin\theta\cos\theta & \cos^2\theta\\
0 & 0 & 0 & -\sqrt{2}\sin\theta\cos\theta & \cos^2\theta-\sin^2\theta & -\sqrt{2}\sin\theta\cos\theta\\
0 & 0 & 0 & \cos^2\theta & -\sqrt{2}\sin\theta\cos\theta & \sin^2\theta
\end{pmatrix}.$%
}
\end{align}
The unitary transformation matrix of a 50/50 beam-splitter corresponds to $U(\pi/4)$. To implement a C-Z gate, we must add a minus sign on $\vert11\rangle$ while keep other logical basis unchanged. This can be done as follows. First, the 50/50 BS makes $\vert11\rangle\rightarrow(\vert02\rangle+\vert20\rangle)/\sqrt{2}$ because of HOM interference \cite{rbi}. Then, two NS gates are used to introduce a $\pi$-phase shift: $\vert02\rangle+\vert20\rangle\rightarrow-\vert02\rangle-\vert20\rangle$. The output photonic state goes through the same beam-splitter again and becomes $-\vert11\rangle$. Other logical basis $\lbrace\vert00\rangle, \vert01\rangle, \vert10\rangle\rbrace$ will not be changed by the NS gate \cite{cia}. Therefore, a C-Z gate is implemented.

Next we consider the effect of the environment including the resonator dissipation, qubit damping and dephasing, and the imperfection of the beam-splitter. We choose initial state $\vert\phi_0\rangle=(\vert00\rangle+\vert01\rangle+\vert10\rangle+\vert11\rangle)/2$, and define the fidelity of the C-Z gate as $F=\vert\langle\phi_{\text{ideal}}\vert\rho_{\text{out}}\vert\phi_{\text{ideal}}\rangle\vert$ with $\vert\phi_{ideal}\rangle=(\vert00\rangle+\vert01\rangle+\vert10\rangle-\vert11\rangle)/2$. According to the master equation (\ref{eq.drhodt}), $F$ reaches 99.89\% when $\omega_{r}/2\pi=\omega_{q}/4\pi=5\:\text{GHz}$,  $\kappa=\gamma=\gamma_{\varphi}=0.05\:\mu\text{s}^{-1}$, and $\theta=\pi/4+0.01$. We have also depicted the input and output state density matrices in Fig.\;\ref{fig.5}, which show our scheme is efficient and robust against decoherence.

\section{Two-photon C-Z gate in the perturbative ultrastrong coupling regime}\label{III}

In this section, we realize the two-photon C-Z gate in the perturbative ultrastrong coupling (p-USC) regime of the two-photon QRM, where $g(\bar{n}+1)\ll\omega_{q}, \omega_{q}+2\omega_{r}$. Here $\bar{n}\leq2$ for arbitrary initial photonic state (\ref{eq.psi0}). In this regime, the two-photon QRM reduces to the two-photon Bloch-Siegert Hamiltonian \cite{tpq}
\begin{align}
H_{2BS}=&H_{2JC}-\omega_{2BS}a^{\dagger}a+(\frac{\omega_{2BS}}{2}+\frac{\Omega_{q}}{2})\sigma_{z}\nonumber\\
&+(\frac{\omega_{2BS}}{2}+2\Omega_{q})\sigma_{z}\big(a^{\dagger}a+(a^{\dagger}a)^{2}\big),
\end{align}
where $\omega_{2BS}=2g^2/(2\omega_{r}+\omega_{q})$ and $\Omega_{q}=2g^2/\omega_{q}$ are the two-photon Bloch-Siegert shifts \cite{sco,pdp}. Choosing
\begin{align}
H_{0}=(\omega_{r}-\omega_{2BS})a^\dagger a+(\frac{\omega_{q}}{2}+\frac{\omega_{2BS}}{2}+\frac{\Omega_{q}}{2})\sigma_{z},
\label{eq.h0}
\end{align}
we obtain
\begin{align}
H_{2BS}^{int}=&(\frac{\omega_{2BS}}{2}+2\Omega_{q})\sigma_{z}(a^{\dagger}a+(a^{\dagger}a)^{2})\nonumber\\
&+g(\sigma_{+}a^{2}e^{i(\omega_{q}-2\omega_{r}+3\omega_{2BS}+\Omega_{q})t}+H.c.)
\end{align}
in the interaction picture. To make it time-independent, we choose $\omega_{q}-2\omega_{r}+3\omega_{2BS}+\Omega_{q}=0$. So that
\begin{eqnarray}
\frac{g}{\omega_{r}}=\frac{\sqrt{4r-r^{3}}}{2\sqrt{1+2r}},
\end{eqnarray}
where $r=\omega_{q}/\omega_{r}$. Now the total excitation number operator $a^\dagger a+2\sigma_+\sigma_-$ is still conserved. For initial state (\ref{eq.psi0}), the involved Hilbert space consists of $\lbrace\vert0,g\rangle,\vert1,g\rangle,\vert0,e\rangle,\vert2,g\rangle\rbrace$, and $H_{2BS}^{int}$ reads
\begin{align}
\begin{pmatrix}
0 & 0 & 0 & 0 \\
0 & B/3 & 0 & 0 \\
0 & 0 & 0 & \sqrt{2}g \\
0 & 0 & \sqrt{2}g & B 
\end{pmatrix},
\end{align}
where $B=-6(\omega_{2BS}/2+2\Omega_{q})$ is a function of $r$. By solving the Schr\"odinger equation analytically, we obtain
\begin{align}
\vert&\psi(t)\rangle=\alpha_{0}\vert0,g\rangle+\alpha_{1}e^{-i\frac{B}{3}t}\vert1,g\rangle\nonumber\\
&+\alpha_{2}e^{-i\frac{B}{2}t}\Big[\frac{-i2\sqrt{2}g\sin(\frac{1}{2}\sqrt{B^{2}+8g^{2}}t)}{\sqrt{B^{2}+8g^{2}}}\vert0,e\rangle\nonumber\\
&+\Big(\cos(\frac{1}{2}\sqrt{B^{2}+8g^{2}}t)-\frac{iB\sin(\frac{1}{2}\sqrt{B^{2}+8g^{2}}t)}{\sqrt{B^{2}+8g^{2}}}\Big)\vert2,g\rangle\Big].
\label{eq.psi2}
\end{align}
\begin{figure}[t]
\centering
\includegraphics[scale=0.3]{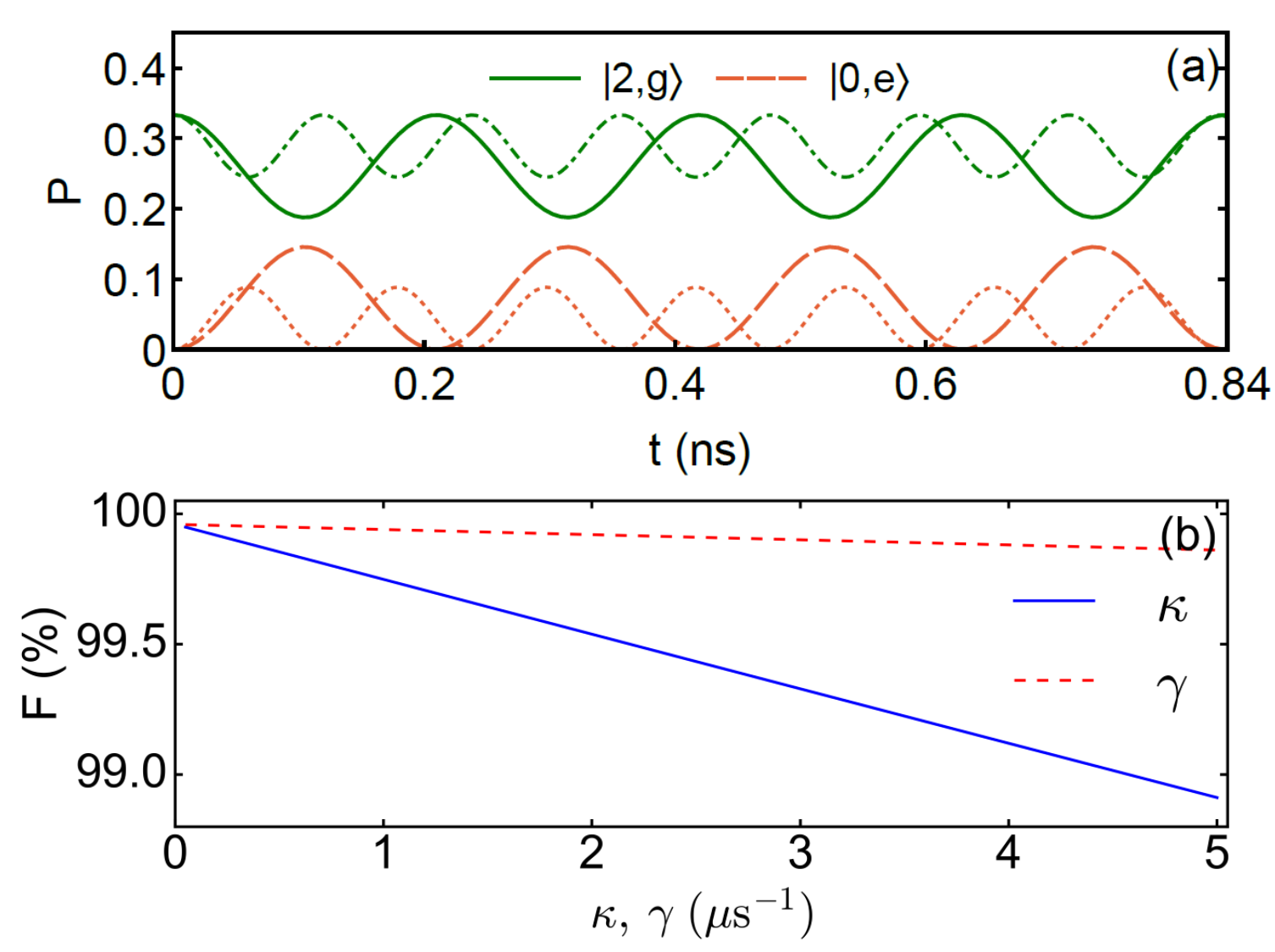}
\renewcommand\figurename{\textbf{FIG.}}
\caption[1]{(a) Dynamical evolution governed by $H_{2BS}^{int}$. $\omega_{r}/2\pi=5\:\text{GHz}$, and $\kappa=\gamma=0.05\:\mu\text{s}^{-1}$. The solid and dashed oscillations correspond to the parameters $\omega_{q}/\omega_{r}=1.87$ and $g/\omega_{r}=0.223$ for $k=4$, while the dot-dashed and dotted oscillations correspond to the parameters $\omega_{q}/\omega_{r}=1.742$ and $g/\omega_{r}=0.306$ for $k=7$. A nonlinear $\pi$-phase shift is realized after $4$ and $7$ oscillations respectively. (b) Effects of the environment on the fidelity of the C-Z gate with the parameters of $k=4$.}
\label{fig.6}
\end{figure}
It can be seen that $\vert0,g\rangle$ is unchanged, and $\vert1,g\rangle$ will acquire a phase of $-Bt/3$. $\vert2,g\rangle$ will oscillate with $\vert0,e\rangle$ and acquire a phase of $-BT/2+\pi$ after the population oscillation of a period $T=2\pi/\sqrt{B^{}+8g^{2}}$ in the two-dimensional subspace $\lbrace\vert2,g\rangle,\vert0,e\rangle\rbrace$ [See Fig. \hyperref[fig.6]{6(a)}]. Unlike the two-photon JC model, here the population can not be completely transferred. Supposing $t=kT$, the acquired phase for $\vert0,g\rangle$, $\vert1,g\rangle$ and $\vert2,g\rangle$ reads
\begin{align}
&\theta_{0}=0,\\
&\theta_{1}=-\frac{B}{3}kT,\\
&\theta_{2}=-\frac{B}{2}kT+k\pi,
\end{align}
respectively. To realize the NS gate, it is required that
\begin{align}
&\theta_{1}-\theta_{0}=2n\pi,\label{eq.kn}\\
&\theta_{2}-\theta_{0}=(2m+1)\pi,\label{eq.km}
\end{align}
where $k, n$ and $m$ are integers. The minimum $t=kT$ satisfying Eqs. (\ref{eq.kn}) and (\ref{eq.km}) reads 
\begin{eqnarray}
kT=6\pi/B,~~~k\text{ being even},\label{eq.keven}\\
kT=12\pi/B,~~~k\text{ being odd}.\label{eq.kodd}
\end{eqnarray}
Since $T$ and $B$ are both functions of $r$, Eqs. \eqref{eq.keven} and \eqref{eq.kodd} can be combined as
\begin{align}
\frac{k}{3+(-1)^{k+1}}&=\frac{(1+2r)}{2(2-r)(8+5r)}\times\nonumber\\
&\sqrt{\frac{(2-r)(1152-r(-880+r(230+209r)))}{(1+2r)^{2}}}.
\label{eq.kr}
\end{align}
Since $r=\omega_{q}/\omega_{r}>0$, $k=4,6,7,8,9,\ldots$ to satisfy Eq. (\ref{eq.kr}). Once $k$ is fixed, we can obtain $\omega_{q}/\omega_{r}$, $g/\omega_{r}$ and gate time $t$ in unit of $\omega_{r}^{-1}$, as shown in Tab.\;\ref{tab.1}. We plot the population of $\vert 2,g\rangle$ and $\vert 0,e\rangle$ for $k=4$ and 7 in Fig.\;\hyperref[fig.6]{6(a)}. Once the NS gate is realized, the C-Z gate can be constructed with the same way as in the last section.

Finally, we study the dissipative dynamics in p-USC regime with dressed-state master equation \cite{dau,pbi}
\begin{align}
\dot{\rho}(t)=&-i[H_{2BS}^{int},\rho(t)]+\nonumber\\
&\sum_{j,k>j}(\Gamma_{\kappa}^{jk}+\Gamma_{\gamma}^{jk})(D(|j\rangle\langle k|)\rho(t)),
\end{align}
where $D(L)\rho=(2L\rho L^{\dagger}-L^{\dagger}L\rho-\rho L^{\dagger}L)/2$, and $\lbrace\vert j\rangle\rbrace_{j=0,1,2\ldots.}$ are eigenstates of the Hamiltonian $H_{2BS}$ with eigenenergy $\epsilon_{j}$. The decay rates read
\begin{align}
&\Gamma_{\kappa}^{jk}=\kappa\frac{\Delta_{kj}}{\omega_{r}}|\langle k|(a+a^\dagger)|j\rangle|^2,\\
&\Gamma_{\gamma}^{jk}=\gamma\frac{\Delta_{kj}}{\omega_{q}}|\langle k|\sigma_{x}|j\rangle|^{2},
\end{align}
where $\Delta_{kj}=\epsilon_{k}-\epsilon_{j}$. The numerical results show the C-Z gate fidelity is robust against decoherence, as depicted in Fig. \hyperref[fig.6]{6(b)}.

Here $H_{0}$ (\ref{eq.h0}) is used in our interaction picture. If we replace it with a commonly used $H_{0}=\omega_{r}a^{\dagger}a+\omega_{q}/2\;\sigma_{z}$, we can also construct the NS gate with the same steps. In this way, the resonance condition is fixed at $\omega_{q}=2\omega_{r}$, and the relationship between $g$ and gate time $t$ can also be obtained by repeating the above derivation,
\begin{align}
&\frac{k}{3+(-1)^{k+1}}=\frac{\sqrt{2}\omega_{r}}{g}\sqrt{1+32\big(\frac{g}{\omega_{r}}\big)^{2}},\\
&t=\frac{k\pi}{\sqrt{2}g\sqrt{1+32(g/\omega_{r})^{2}}}.~~~\big(g(\bar{n}+1)\ll\omega_{q}\big)
\end{align}
However, the fastest gate time is about 3.6 ns with $g=0.236\omega_r$ and $k=20$, which is longer than that in the previous result.

\begin{table}[t]
\centering
\begin{tabularx}{0.48\textwidth}{XXXXXX}
\hline
$k$ & $r$ & \begin{tabular}[c]{@{}c@{}}$\omega_{q}/(2\pi)$\\(GHz)\end{tabular} & \begin{tabular}[c]{@{}c@{}}$g/(2\pi)$\\(GHz)\end{tabular} & \begin{tabular}[c]{@{}c@{}}$t$\\(ns)\end{tabular} & \begin{tabular}[c]{@{}c@{}}$F$\\(\%)\end{tabular} \\
\hline
4 & 1.870 & 9.35 & 1.115 & 0.84 & 99.95 \\
6 & 1.964 & 9.82 & 0.595 & 3.1 & 99.90 \\
7 & 1.742 & 8.71 & 1.53 & 0.83 & 99.95 \\
8 & 1.982 & 9.91 & 0.42 & 6.2 & 99.83 \\
9 & 1.916 & 9.58 & 0.9 & 2.6 & 99.92 \\
\hline
\end{tabularx}
\renewcommand{\thetable}{\arabic{table}}
\renewcommand\tablename{\textbf{TAB.}}
\caption[1]{Parameters correspond to different solutions $k$ and $r$. For each set of parameters, the fidelity and gate time of the C-Z gate are also provided. $\omega_{r}/2\pi$=5\:\text{GHz}, $\kappa=\gamma=\gamma_{\varphi}=0.05\:\mu\text{s}^{-1}$ and the beam-splitter $\theta=\pi/4+0.01$.}
\label{tab.1}
\end{table}

\section{Two-photon C-Z gate in the dispersive regime}
\label{IV}

Finally, we propose a scheme to realize the C-Z gate with the two-photon QRM in the dispersive regime. When the frequencies of the resonator mode and the qubit are far detuned, the system can be described by a dispersive Hamiltonian, which is widely used for qubit readout in circuit QED \cite{sqr,moo}. The two-photon QRM in the dispersive regime, i.e., $\vert\delta\vert=\vert\omega_{q}-2\omega_{r}\vert\gg g(\bar n+1)$, is given by \cite{tpq}
\begin{align}
H_{2JC}^{dis}=&(\omega_{r}+\chi)a^{\dagger}a+(\omega_{q}+\chi)\frac{\sigma_{z}}{2}\nonumber\\
&+\frac{\chi}{2}\sigma_{z}\big(a^{\dagger}a+(a^{\dagger}a)^{2}\big),
\label{eq.H2JCdis}
\end{align}
with $\chi=2g^2/\vert\delta\vert$. The last term in Eq. (\ref{eq.H2JCdis}) consists of a nonlinear Kerr term that depends on the qubit state. Now the Hamiltonian is much simplified and one can easily obtain
\begin{align}
\vert\psi(t)\rangle&=\alpha_{0}e^{-iE_{0}t}\vert0,g\rangle+\alpha_{1}e^{-iE_{1}t}\vert1,g\rangle+\alpha_{2}e^{-iE_{2}t}\vert2,g\rangle\nonumber\\
&=\alpha_{0}e^{i\frac{1}{2}(\omega_{q}+\chi)t}\vert0,g\rangle+\alpha_{1}e^{i(\frac{1}{2}(\omega_{q}+\chi)-\omega_{r})t}\vert1,g\rangle\nonumber\\
&\quad+\alpha_{2}e^{i(\frac{1}{2}(\omega_{q}+3\chi)-2\omega_{r})t}\vert2,g\rangle,
\end{align}
\begin{figure}[t]
\centering
\includegraphics[scale=0.28]{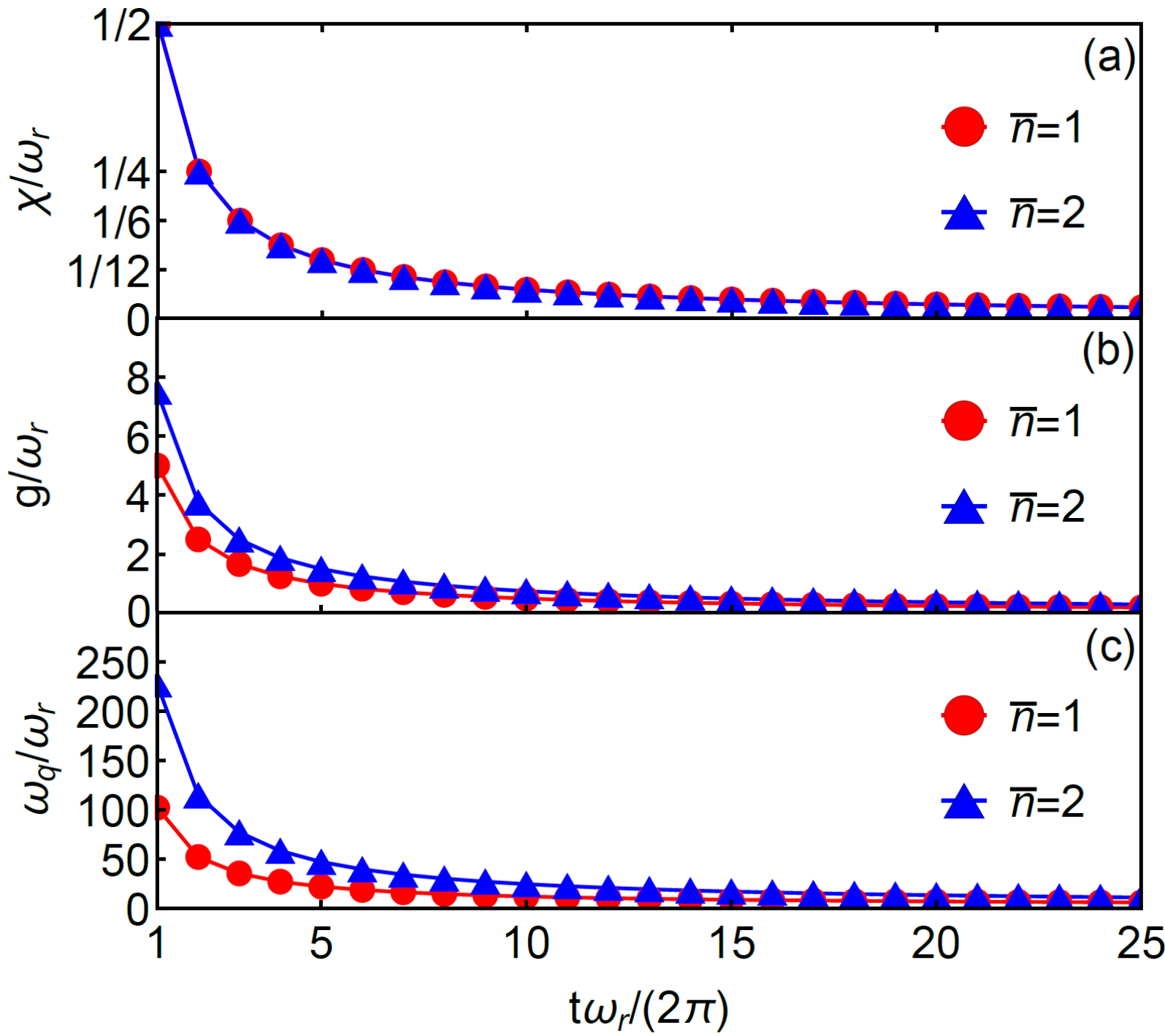}
\renewcommand\figurename{\textbf{FIG.}}
\caption[1]{(a)(b)(c) Relationship between C-Z gate time $t$ and $\chi$, $g$, and $\omega_{q}$.}
\label{fig.7}
\end{figure}
with initial state (\ref{eq.psi0}). To realize the nonlinear $\pi$-phase shift of the NS gate, the gate time $t$ satisfies
\begin{align}
&\Delta E_{01}t=(E_{1}-E_{0})t=\omega_{r}t=2n\pi,\\
&\Delta E_{12}t=(E_{2}-E_{1})t=(\omega_{r}-\chi)t=(2m-1)\pi,\label{cphi}\\
&\frac{\Delta E_{01}}{\Delta E_{12}}=\frac{\omega_{r}}{\omega_{r}-\chi}=\frac{2n}{2m-1}, \label{eq.chinm}
\end{align}
where $n$ and $m$ are integers. Since $\chi$ is normally smaller than $\omega_r$ in the dispersive regime, $2n>2m-1$. $\chi/\omega_r=(2n-2m+1)/2n$. To make parameters more practical in experiment, we chose $n=m$ to minimize $\chi/\omega_r=1/2n$ as plotted in Fig.\;\hyperref[fig.7]{7(a)}. Considering $t=2n\pi/\omega_r$, we obtain
\begin{align}
&\frac{g}{\omega_{r}}=\frac{(\bar{n}+1)\pi}{2t\omega_{r}}\frac{|\delta|}{g(\bar{n}+1)},\\
&\frac{\omega_{q}}{\omega_{r}}=\frac{5(\bar{n}+1)^2\pi}{t\omega_{r}}\frac{|\delta|}{g(\bar{n}+1)}+2.
\end{align}
\\
Since $|\delta|/\big(g(\bar n+1)\big)\gg1$, we chose it as $10$ to depict the relationship between the gate time and $g/\omega_r$ for $\bar n=1$ and 2 in Fig. \hyperref[fig.7]{7(b)}. It can be seen that the fastest time reaches $2\pi/\omega_r$. However, a large $g/\omega_r=5$ is required. The gate time reaches $20\pi/\omega_r$ when $g=0.5\omega_r$. Meanwhile, since $|\delta|/\big(g(\bar n+1)\big)$ is chosen as $10$, we obtain $\omega_q/\omega_r$, and depicted it in Fig. \hyperref[fig.7]{7(c)} for $\bar n=1$ and 2. One can choose suitable $g$, $\omega_r$ and $\omega_q$ in experiment to realize the C-Z gate with operation time shown in Fig. \ref{fig.7}.

\begin{figure}[t]
\centering
\includegraphics[scale=0.26]{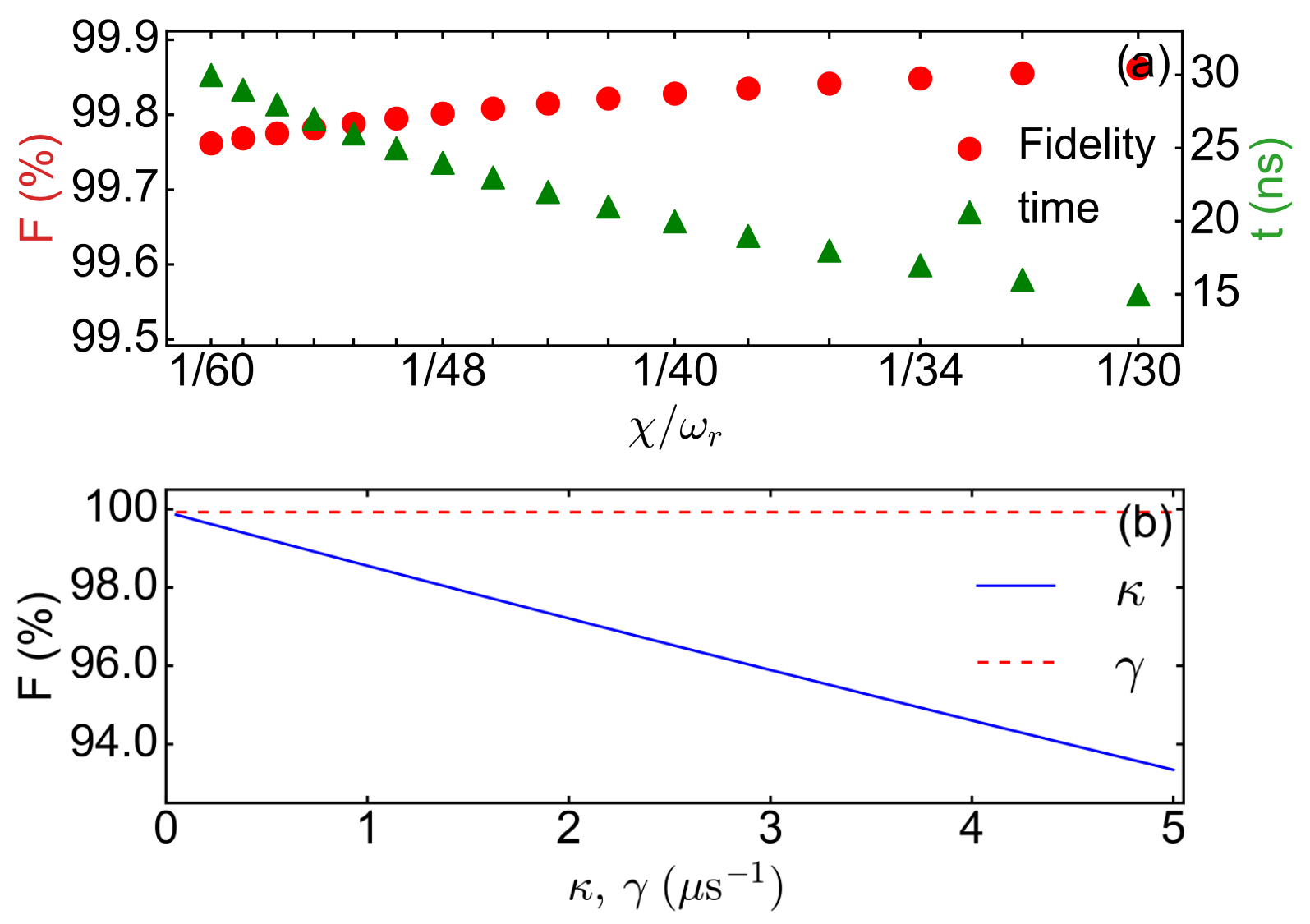}
\renewcommand\figure{\textbf{FIG.}}
\caption[1]{(a) Fidelity and operation time of the C-Z gate. Simulation fidelity: red dots; gate time: green triangles. $\omega_{r}/2\pi=1\:\text{GHz}$, $\omega_{q}/2\pi=12\:\text{GHz}$. The environment dissipation parameters $\gamma=\kappa=0.01\:\mu\text{s}^{-1}$, and the beam-splitter $\theta=\pi/4+0.01$. (b) Environment effects on the fidelity of the C-Z gate. $\omega_{r}/2\pi=1\:\text{GHz},\;\omega_{q}/2\pi=12\:\text{GHz}, g/2\pi=0.395\:\text{GHz}$.}
\label{fig.8}
\end{figure}

Considering realistic experimental conditions, we choose $\omega_{r}/2\pi=1\:\text{GHz}$ and $\vert\delta\vert=\vert\omega_{q}-2\omega_{r}\vert=10\omega_{r}$. Therefore, $g=\sqrt{\vert\delta\vert\omega_{r}/4n}$, taking into account that $\chi/\omega_{r}=1/2n$. The resonator dissipation rate and qubit damping rate are chosen as $\kappa=\gamma=0.01\:\mu\text{s}^{-1}$. We have also considered the imperfection of the beam-splitter by choosing the angle $\theta=\pi/4+0.01$. Then, we show the relationship between gate time $t$, fidelity $F$ and $\chi/\omega_r$ in Fig. \hyperref[fig.8]{8(a)}, by using the dressed master equation (\ref{eq.drhodt}). Finally, we choose one of the appropriate parameters in Fig. \hyperref[fig.8]{8(a)} ($\chi/\omega_{r}=1/36$, $\vert\delta\vert/\omega_{r}=10$, $g/\omega_{r}=0.395$), to consider the effect of the environment. Interestingly, the fidelity decreases with $\kappa$ but will not change with $\gamma$, as depicted in Fig. \hyperref[fig.8]{8(b)}. This is because the qubit is set to the ground state initially and there is not qubit transition  terms in $H_{2JC}^{dis}$ (\ref{eq.H2JCdis}). Therefore it will always stay in the ground state, thus avoiding the effect of damping.

An advantage of dispersive coupling is that it can be used to implement the C-phase gate directly. It does not cause population oscillation, and add an arbitrary phase before $\vert 2,g\rangle$. A C-phase gate can be realized if we replace $(2m-1)\pi$ with an arbitrary $\phi$ in Eq. \eqref{cphi}.

\section{Input and output of the photonic states}
So far we only consider the interaction between the qubit and photon in the circuit QED system, however, the input and out put of the photonic states must be included in the whole experiment process. First, we must transfer the photonic state $\vert \phi\rangle$ like $\alpha_0\vert0\rangle_w+\alpha_1\vert1\rangle_w+\alpha_2\vert2\rangle_w$ in the waveguide to the circuit. Normally, $\vert \phi\rangle$ is not an ideal plane wave, but a Lorentz wave packet centered at $\omega_0$ with finite bandwidth $\epsilon$ \cite{liao1,liao2,tanjin}, therefore
\begin{align}
&|1\rangle_w=\int_0^\infty d\omega\mathcal{A}\:\frac{1}{\omega-\omega_{0}-i\epsilon}b_\omega^\dagger|0\rangle ,\\
&\vert 2\rangle_w=\int _0^\infty d\omega\int _0^\omega d\omega'\mathcal{B}\:\left(\frac{1}{\omega-\omega_{0}+i\epsilon}\frac{1}{\omega^{\prime}-\omega'_{0}+i\epsilon}\right.\nonumber\\
&\hspace{3em}  \left.+\frac{1}{\omega'-\omega_{0}+i\epsilon}\frac{1}{\omega-\omega'_{0}+i\epsilon}\right)b_\omega^\dagger b_{\omega'}^\dagger|0\rangle ,
\end{align}
where $b_\omega^\dag$ and $b_{\omega'}^\dag$ are creation operators of the waveguide mode with frequency $\omega$ and $\omega'$ respectively. $\mathcal{A}$ and $\mathcal{B}$ are normalizing constants.
$\vert 2\rangle_w$ takes this form to satisfy the permutation symmetry of the bosons.  For simplicity, we consider two photons have the same center frequency resonant with the resonator $\omega_0=\omega'_0=\omega_r$. 

The input and output process is as follows. First we turn off the resonator-qubit coupling to input $\vert \phi\rangle$ to the resonator, so that the Hamiltonian reads
\begin{equation}\small
H = \omega_{r}a^{\dagger}a  + \int_{0}^{\infty} d\omega \omega b_{\omega}^{\dagger}b_\omega + g'_{wr}\int_{0}^{\infty} d\omega(a^{\dagger}b_{\omega} + b_{\omega}^{\dagger}a),
\end{equation}
where $ g'_{wr}$ is the coupling between the waveguide and resonator. In the interaction picture with respect to $H_0= \omega_{r}(a^{\dagger}a  +  \int_0^\infty d\omega b_{\omega}^{\dagger}b_\omega)$, the interaction Hamiltonian reads
\begin{equation}
H_I=\int_{0}^{\infty} d\omega \Delta_\omega b_{\omega}^{\dagger}b_\omega + g'_{wr}\int_{0}^{\infty} d\omega(a^{\dagger}b_{\omega} + b_{\omega}^{\dagger}a).
\end{equation}
where $\Delta_\omega=\omega-\omega_r$. The excitation number is conserved, so that 
\begin{align}
|\psi(t)\rangle = &A(t)|2\rangle_{r}|0\rangle_w + \int_{0}^{\infty} d\omega B_{\omega}(t)|1\rangle_{r}|1_{\omega}\rangle\nonumber\\& + \int_{0}^{\infty} d\omega \int_{0}^{\omega} d\omega' C_{\omega,\omega'}(t)|0\rangle_{r}|1_{\omega},1_{\omega'}\rangle,
\end{align}
with $|\psi(0)\rangle=|0\rangle_r|\phi\rangle$, where $\vert i\rangle_r$ denotes $i$ photons in the resonator. We aim to obtain $|\psi(t_{in})\rangle =|0\rangle_w(\alpha_0\vert0\rangle_r+\alpha_1\vert1\rangle_r+\alpha_2\vert2\rangle_r)$, however, it seems impossible to transfer all the population in a multimode state $|\phi\rangle$ into a single mode state with a constant coupling. Therefore, we introduce a variable coupler proposed by Yin et al. \cite{car} which connects the waveguide and the resonator. Initially, their coupling $g_{wr}$ is large, so that the population is quickly transferred from the waveguide to the resonator. Then we gradually decrease $g_{wr}$ close to $0$, to prevent the accumulated population in the resonator goes back to the waveguide. By optimizing $g_{wr}$, we can input the waveguide photonic state $|\phi\rangle$ to the resonator. Then we turn on the qubit-resonator coupling $g_{rq}$ to add a $\pi$-phase shift to the two-photon part. Finally, we turn on $g_{wr}$ again to output the photonic state to complete the NS gate. We need to optimize $g_{wr}$ again to ensure the output photon waveform keeps the same as the input one, except for a $\pi$-phase shift.

\begin{figure}[t]
\centering
\includegraphics[scale=0.45]{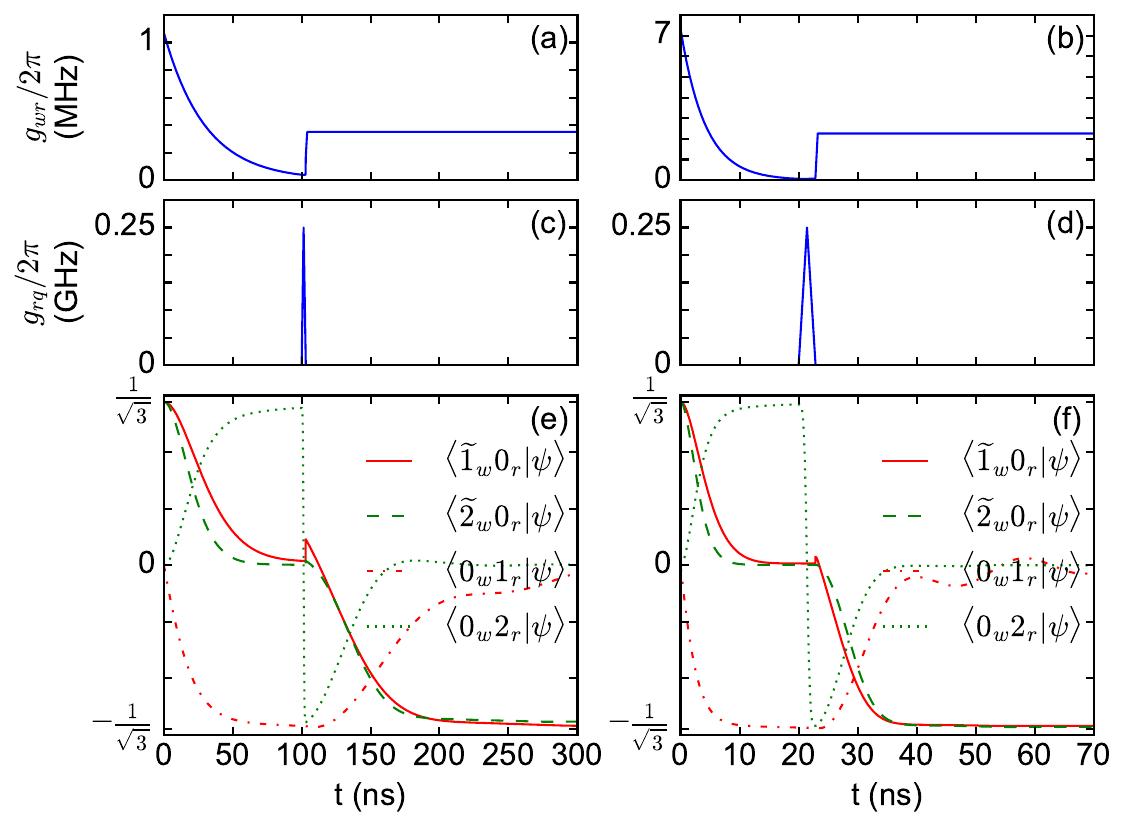}
\renewcommand\figurename{\textbf{FIG.}}
\caption[1]{Numerical simulation of the input and output process of $\vert\phi\rangle=\frac{1}{\sqrt{3}}(|0\rangle_w+|1\rangle_w+|2\rangle_w)$ as a Lorentz wavepacket with width $0.02$ GHz (left panel) and $0.15$ GHz (right panel). The parameters are optimized to ensure the success of this process, and shown in (a), (b), (c) and (d). We project instantaneous state $\vert \psi(t)\rangle$ into  $| \widetilde{1}\rangle_w|0\rangle_r$, $| \widetilde{2}\rangle_w|0\rangle_r$, $| 0\rangle_w|1\rangle_r$ and $| 0\rangle_w|2\rangle_r$ to show the input and output process in (e) and (f). These projections are real because the imaginary parts are vanishing small.}
\label{fig.9}
\end{figure}
Next we testify our proposal by numerical simulations. First we discrete the waveguide modes by introducing a finite but small frequency interval $\delta \omega=k\epsilon/N$ between two adjacent
 modes $\vert 1_{\omega_{m}}\rangle_w$ and $\vert 1_{\omega_{m+1}}\rangle_w$ \cite{htq}, such that the interaction Hamiltonian becomes
\begin{equation}
H_I=\sum_{m}  \Delta_{\omega_m} b_{\omega_m}^{\dagger}b_{\omega_m} + g_{wr}\sum_m (a^{\dagger}b_{\omega_m} + b_{\omega_m}^{\dagger}a).
\end{equation}
We chose $N=100$, and the center frequency equalling $\omega_r$. Then we solve the Schr\"{o}dinger equation to obtain $\vert \psi\rangle$ and optimize $g_{wr}$ accordingly. We consider two kinds of input Lorentz wavepackets with $\alpha_0=\alpha_1=\alpha_2=1/\sqrt{3}$: 1. $\epsilon=0.02$ GHz, $k=5$. 2. $\epsilon=0.15$ GHz, $k=4$. The numerical results are shown in the left and right panel of Fig. \ref{fig.9} respectively, where we define
 \begin{align}
|\widetilde{i}\rangle_w=
\begin{cases}
\exp(-iH_{I0}t)  |i\rangle_w & t<t_{in}+t_q,\\
\exp(-iH_{I0}(t-t_{in}-t_q)) |i\rangle_w& t\geq t_{in}+t_q\\
\end{cases}
 \end{align}    
for $i=1,2$
 to include the  free evolution
 of the photonic states in the waveguide with $H_{I0}=\sum_{m}  \Delta_{\omega_m} b_{\omega_m}^{\dagger}b_{\omega_m}$. 
 $t_q$ is the qubit-resonator interaction time. The population of the waveguide modes almost vanishes at $t_{in}+t_q$, so that the accumulated dynamical phase is eliminated and $\exp(-iH_{I0}t)$ restart action when $g_{wr}$ is turned on again at $t=t_{in}+t_q$. We project $\vert \psi\rangle$ into $| \widetilde{1}\rangle_w|0\rangle_r$, $| \widetilde{2}\rangle_w|0\rangle_r$, $| 0\rangle_w|1\rangle_r$ and $| 0\rangle_w|2\rangle_r$ to show the input and output process in Figs. \ref{fig.9}  (e) and (f). The imaginary part is vanishing small, so that these projections are real. 
 
\begin{figure}[t]
\centering
\includegraphics[scale=0.43]{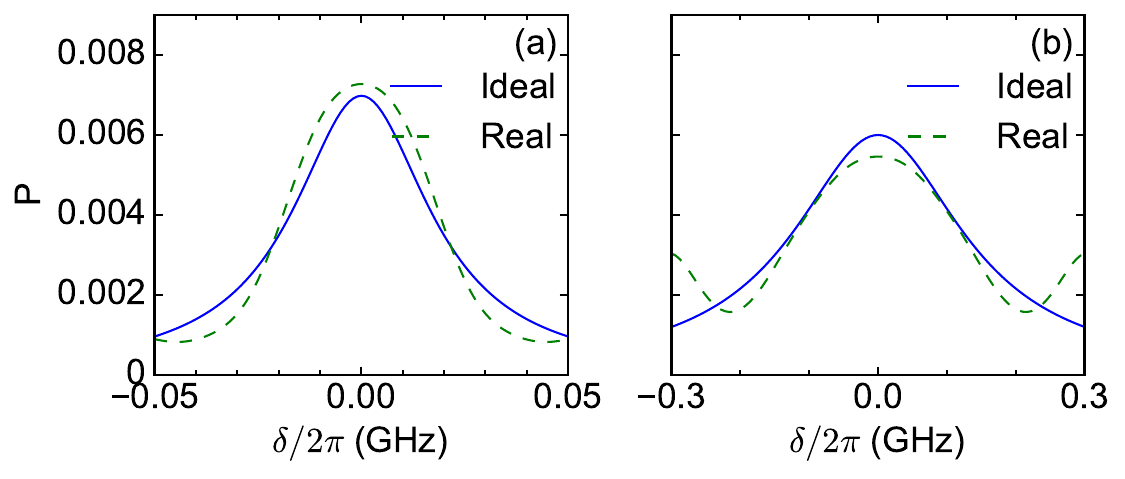}
\renewcommand\figurename{\textbf{FIG.}}
\caption[1]{Waveform of the input (ideal) and output (real) single-photon part of the photonic state in the waveguide. (a) $\epsilon=0.02$ GHz. (b) $\epsilon=0.15$ GHz.}
\label{fig.10}
\end{figure}

When $\epsilon=0.02$ GHz, $\vert \phi\rangle\vert 0\rangle_r$ is transferred to $\frac{1}{\sqrt{3}}|0\rangle_w(|0\rangle_r-|1\rangle_r+|2\rangle_r)$ in $t_{in}=100$ ns, as shown in Fig. \ref{fig.9} (e). The minus sign before $|0\rangle_w|1\rangle_r$ coincides with the single-photon resonant scattering result in waveguide QED \cite{liao1,liao2,tanjin}, which can be eliminated by a linear phase shifter. Or else, one can interchange the definition of $\vert 0\rangle$ and $\vert 1\rangle$, then the C-Z gate can be realized subject to a trivial global phase $\pi$ \cite{tcg}.
$g_{wr}/2\pi=1.07\exp(-0.0333t)$ MHz, as depicted in Fig. \ref{fig.9} (a). $g_{rq}$ is then turned on as  
\begin{align}
g_{rq}(t)=&
\begin{cases}
2g_{0}(t-t_{in})/t_{q}  &t_{in}< t<t_{in}+t_{q}/2,\\
-2g_{0}(t-t_{in}-t_{q})/t_{q}  &t_{in}+t_{q}/2\le t\le t_{in}+t_{q},\\
0  &t<t_{in}\,\&\,t>t_{in}+t_{q},
\end{cases}
\end{align}
where $g_0/2\pi=0.25$ GHz, as depicted in Fig. \ref{fig.9} (c).
We consider the strong qubit-resonator coupling regime, where the wavefunction of the circuit QED system reads
\begin{align}
\vert\psi(t)\rangle=&\cos\big(\sqrt{2}\int_{0}^{t}g_{rq}(t^{\prime})\,dt^{\prime}\big)\vert2,g\rangle\nonumber\\
&-i\sin\big(\sqrt{2}\int_{0}^{t}g_{rq}(t^{\prime})\,dt^{\prime}\big)\vert0,e\rangle.
\end{align} 
Since $g_{rq}(t)$ is linear, a $\pi$-phase is accumulated for the two-photon part in $t_q=\sqrt{2}\pi/g_0=2.8$ ns. When $t>t_{in}+t_q$, $g_{rq}$ is turned off and $g_{wr}/2\pi$ is linearly increase to $0.35$ MHz to release the $\alpha_0\vert0\rangle_r-\alpha_1\vert1\rangle_r-\alpha_2\vert2\rangle_r$ to the waveguide. Since we have optimized $g_{wr}$, the output photonic waveform is almost the same as the input one. We plot the input and output single-photon part waveform in Fig. \ref{fig.10} (a), whose overlap reaches $99.3\%$.
 \begin{figure}[t]
\centering
\includegraphics[scale=0.38]{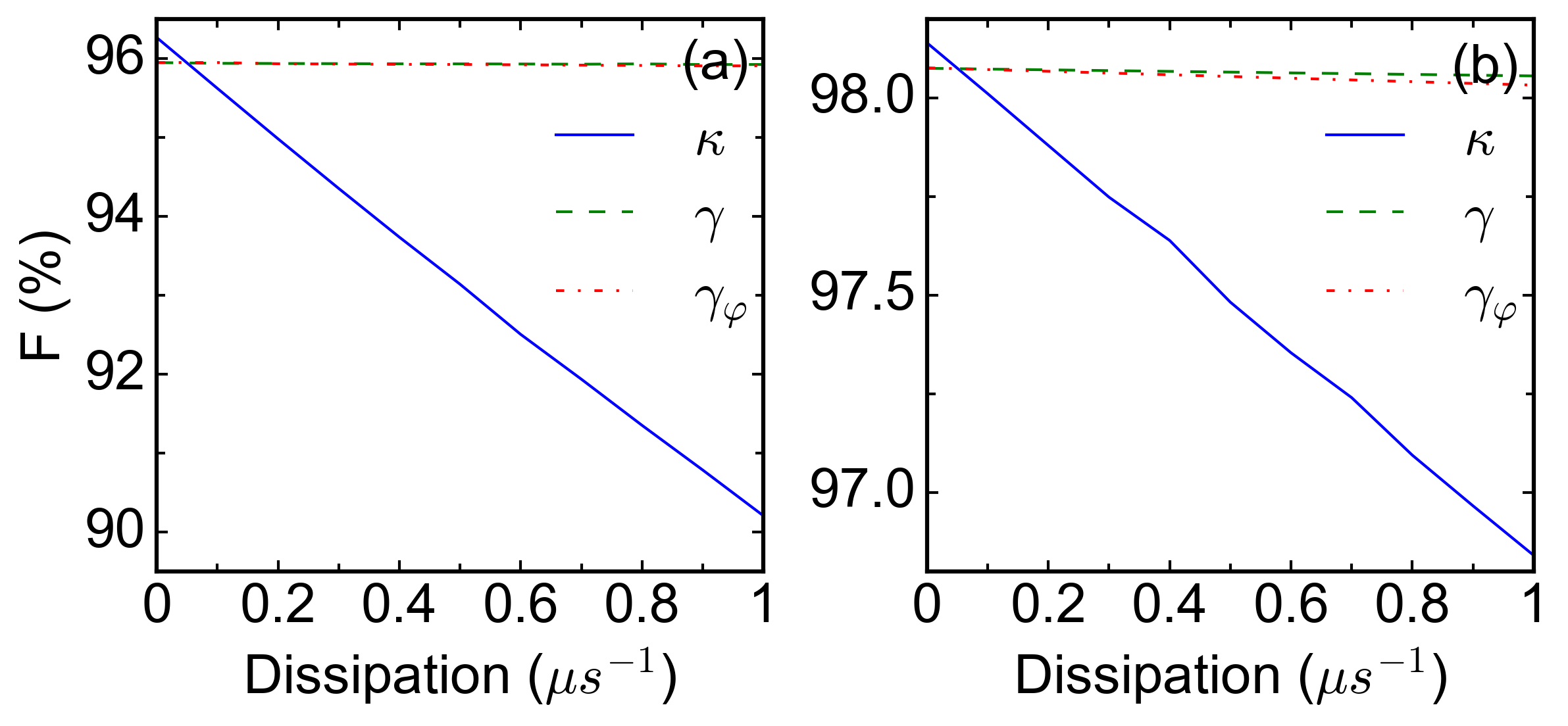}
\renewcommand\figurename{\textbf{FIG.}}
\caption[1]{The fidelity of the NS gate during the whole input and output process as a function of resonator dissipation $\kappa$ ($\gamma$ and $\gamma_{\phi}$ fixed at $0.05$ $\mu s^{-1}$), qubit damping $\gamma$ ($\kappa$ and $\gamma_{\phi}$ fixed at $0.05$ $\mu s^{-1}$) and dephasing $\gamma_{\phi}$ ($\gamma$ and $\kappa$ fixed at $0.05$ $\mu s^{-1}$). (a) Input state $\frac{1}{\sqrt{3}}(|0\rangle_w+|1\rangle_w+|2\rangle_w)$ consists of Lorentz wavepackets with bandwidth $0.02$ GHz. (b) Input state $\frac{1}{\sqrt{3}}(|0\rangle_w+|1\rangle_w+|2\rangle_w)$ consists of Lorentz wavepackets with bandwidth $0.15$ GHz. }
\label{fig.11}
\end{figure}
 
When the Lorentz wavepacket bandwidth reaches $0.15$ GHz, the whole input and output process can be done in $70$ ns, because a larger bandwidth indicates a stronger waveguide-resonator interaction, as shown in the right panel of Fig. \ref{fig.9}. $g_{wr}/2\pi=7.3\exp(-0.24t)$ MHz, and $\vert \phi\rangle$ can be transferred into the resonator in $t_{in}=20$ ns. After interaction with the qubit in  $t_q=\sqrt{2}\pi/g_0=2.8$ ns, we increase $g_{wr}/2\pi$ linearly to $2.26$ MHz. Finally, the photonic state is released to the waveguide with almost the same waveform as $\vert \phi\rangle$. The single-photon part input and output waveform is shown in Fig. \ref{fig.10} (b), whose overlap reaches $98.5\%$.

Finally, we consider the decoherence effects during the whole process. Now the interaction time is much longer than previous sections, so the the effect of environment is more remarkable. We define fidelity $F=\vert\langle\psi_{\text{ideal}}\vert\rho_{\text{out}}\vert\psi_{\text{ideal}}\rangle\vert$ with $\vert\psi_{\text{ideal}}\rangle=\frac{1}{\sqrt{3}}|0\rangle_r(|0\rangle_w-|\widetilde{1}\rangle_w-|\widetilde{2}\rangle_w)$ and plot its relation with decoherence in Fig. \ref{fig.11}. It is more sensitive to $\kappa$ than to $\gamma$ since the SQUID
 is populated for the longest duration, considering the transfers from and to the TL. Now one has to deal with a high dimensional ($\approx N^2/2$) density matrix in the master equation, so one may reduce $N$ if the computer resource is limited.

\section{Conclusion}\label{V}
We have presented a deterministic scheme for the two-photon nonlinear sign gate using variants of the two-photon QRM, which can be implemented in circuit QED. Initially, we consider the strong coupling regime. The nonlinearity is introduced through resonant interactions between microwave photons and a superconducting qubit, taking the form of a second-order nonlinear coupling. Subsequently, we construct a deterministic two-photon C-Z gate using the KLM scheme, which is robust against decoherence sources under current experimental conditions. We also achieve the C-Z gate in the p-USC regime and C-phase gate in the large detuning regime with specific parameters by analyzing the dynamics, which operates fast and remains robust despite imperfect experimental conditions. Finally, we show that the photonic state in the waveguide can be transferred into the resonator through a variable coupler, and released with almost the same waveform after interaction with the qubit, with a $\pi$-phase shift implemented.
Our results offer a novel approach to realizing deterministic two-photon C-Z gates with the two-photon QRM and are expected to contribute to the advancement of quantum computing with microwave photons.

\begin{acknowledgments}
The authors thank Jie-Qiao Liao, Zhi-hui Peng, Xun-wei Xu and Jin-Lei Tan for their helpful discussions. 
 L. L. acknowledges the support from grants PID2022-136228NB-C21 and PID2022-136228NB-C22 funded by MCIN/AEI/10.13039/50110001103 and “ERDF A way of making Europe”, the Ministry for Digital Transformation and of Civil Service of the Spanish Government through the QUANTUM ENIA project call - Quantum Spain project, and by the European Union through the Recovery, Transformation and Resilience Plan - NextGenerationEU within the framework of the “Digital Spain 2026 Agenda”.
\end{acknowledgments}

\end{document}